\documentclass[]{nature}
\makeatletter\if@twocolumn\PassOptionsToPackage{switch}{lineno}\else\fi\makeatother

\usepackage{tabulary,graphicx,amsmath,amsfonts,amssymb}
\usepackage[utf8x]{inputenc}
\usepackage[T1]{fontenc}

\makeatletter
\renewenvironment{figure}
               {\@float{figure}}
               {\end@float}
\renewenvironment{figure*}
               {\@dblfloat{figure}}
               {\end@dblfloat}

\renewenvironment{table*}
               {\@dblfloat{table}}
               {\end@dblfloat}

\makeatother

\usepackage{url,multirow,morefloats,floatflt,cancel,tfrupee}
\makeatletter

\AtBeginDocument{\@ifpackageloaded{textcomp}{}{\usepackage{textcomp}}}
\makeatother
\usepackage{colortbl}
\usepackage{xcolor}
\usepackage{pifont}
\usepackage[nointegrals]{wasysym}
\makeatletter

\def\mcWidth#1{\csname TY@F#1\endcsname+\tabcolsep}

\def\cAlignHack{\rightskip\@flushglue\leftskip\@flushglue\parindent\z@\parfillskip\z@skip}
\def\rAlignHack{\rightskip\z@skip\leftskip\@flushglue \parindent\z@\parfillskip\z@skip}

\@ifundefined{etal}{}{}

\usepackage{ifxetex}
\ifxetex\else\if@twocolumn\@ifpackageloaded{stfloats}{}{\usepackage{dblfloatfix}}\fi\fi

\AtBeginDocument{
\expandafter\ifx\csname eqalign\endcsname\relax
\def\eqalign#1{\null\vcenter{\def\\{\cr}\openup\jot\m@th
  \ialign{\strut$\displaystyle{##}$\hfil&$\displaystyle{{}##}$\hfil
      \crcr#1\crcr}}\,}
\fi
}

\AtBeginDocument{%
  \@ifpackageloaded{endfloat}%
   {\renewcommand\efloat@iwrite[1]{\immediate\expandafter\protected@write\csname efloat@post#1\endcsname{}}}{\newif\ifefloat@tables}%
}%

\def\BreakURLText#1{\@tfor\brk@tempa:=#1\do{\brk@tempa\hskip0pt}}
\let\lt=<
\let\gt=>
\def\processVert{\ifmmode|\else\textbar\fi}

\@ifundefined{subparagraph}{
\def\subparagraph{\@startsection{paragraph}{5}{2\parindent}{0ex plus 0.1ex minus 0.1ex}%
{0ex}{\normalfont\small\itshape}}%
}{}

\newcommand\role[1]{\unskip}
\newcommand\aucollab[1]{\unskip}
  
\@ifundefined{tsGraphicsScaleX}{\gdef\tsGraphicsScaleX{1}}{}
\@ifundefined{tsGraphicsScaleY}{\gdef\tsGraphicsScaleY{.9}}{}
\def\checkGraphicsWidth{\ifdim\Gin@nat@width>\linewidth
	\tsGraphicsScaleX\linewidth\else\Gin@nat@width\fi}

\def\checkGraphicsHeight{\ifdim\Gin@nat@height>.9\textheight
	\tsGraphicsScaleY\textheight\else\Gin@nat@height\fi}

\def\fixFloatSize#1{}
\let\ts@includegraphics\includegraphics

\def\inlinegraphic[#1]#2{{\edef\@tempa{#1}\edef\baseline@shift{\ifx\@tempa\@empty0\else#1\fi}\edef\tempZ{\the\numexpr(\numexpr(\baseline@shift*\f@size/100))}\protect\raisebox{\tempZ pt}{\ts@includegraphics{#2}}}}

\AtBeginDocument{\def\includegraphics{\@ifnextchar[{\ts@includegraphics}{\ts@includegraphics[width=\checkGraphicsWidth,height=\checkGraphicsHeight,keepaspectratio]}}}

\DeclareMathAlphabet{\mathpzc}{OT1}{pzc}{m}{it}

\def\URL#1#2{\@ifundefined{href}{#2}{\href{#1}{#2}}}

\def\UrlOrds{\do\*\do\-\do\~\do\'\do\"\do\-}%
\g@addto@macro{\UrlBreaks}{\UrlOrds}

\edef\fntEncoding{\f@encoding}

\makeatother

\newif\ifmultipleabstract\multipleabstractfalse%
\usepackage[nomarkers,tablesfirst,nolists]{endfloat}

\def\fixFloatSize#1{}
\begin{document}

\title{Cylindrical vector beams reveal radiationless anapole condition in a resonant state}

\maketitle


 Yudong Lu \ensuremath{^{1}}, Yi Xu \ensuremath{^{1,2,}}*, Xu Ouyang \ensuremath{^{1,2}}, Mingcong Xian \ensuremath{^{2}}, Yaoyu Cao \ensuremath{^{1}},Kai Chen \ensuremath{^{1}} , and Xiangping Li \ensuremath{^{1,}}*\ensuremath{^{}}

\ensuremath{^{\textit{1}}} \textit{ Guangdong Provincial Key Laboratory of Optical Fiber Sensing and Communications, Institute of Photonics Technology, Jinan University, Guangzhou, 510632, China}

\ensuremath{^{2}}\textit{Department of Electronic Engineering, College of Information Science and Technology, Jinan University, Guangzhou, 510632, China}

*Corresponding author: yi.xu@osamember.org, xiangpingli@jnu.edu.cn

\begin{abstract}
Nonscattering optical anapole condition is corresponding to the excitation of radiationless field distributions in open resonators, which offers new degrees of freedom for tailoring light-matter interaction. Conventional mechanisms for achieving such a condition relies on sophisticated manipulation of electromagnetic multipolar moments of all orders to guarantee superpositions of vanished moment strengths at the same wavelength. In contrast, here we report on the excitation of optical radiationless anapole hidden in a resonant state of a Si nanoparticle utilizing tightly focused radially polarized (RP) beam. The coexistence of magnetic resonant state and anapole condition at the same wavelength further enables the triggering of resonant state by tightly focused azimuthally polarized (AP) beam whose corresponding electric multipole coefficient could be zero. As a result, high contrast inter-transition between radiationless anapole condition and ideal magnetic resonant scattering can be achieved experimentally in visible spectrum. The proposed mechanism is general which can be realized in different types of nanostructures. Our results showcase that the unique combination of structured light and structured Mie resonances could provide new degrees of freedom for tailoring light-matter interaction, which might shed new light on functional meta-optics.
\end{abstract}\def\keywordstitle{Keywords}


\section{Introduction}

Optical scattering of a nanopaticle under the excitation of a plane wave is usually determined by its predominant electromagnetic multipole moment \cite{book}. Such predominant multipole moment can even decide the electric or magnetic nature of the scattering in all-dielectric photonics \cite{ML}. Therefore, tailoring such predominant electromangic multipole moment becomes an effective and unified way to manipulate optical scatteirng \cite{ABO}. Many abnormal optical scattering phenomena have been proposed to enable new possibility and functionality in photonics \cite{ABO}. The nonscattering electromagnetic state is one of the typical examples \cite{anapole1,anapole2,anapole3,high_td,anapole_nl1,S_anapole1,anapole4,anapole5,anapole6,anapole_nl2,EM_anapole,anapole7,S_anapole2,S_anapole3,anapole8,anapole_LN}  [see these elaborated reviews in Refs. \cite{R1,R2,R3,R4} for details]. In contrast to the embedded eigenstates or bound state in the continuum which cannot be accessed by the excitation in the continuum, the so-called electromagnetic anapole condition provides a nonscattering condition sustained under the excitation of external field\cite{5,24}, which resembles a promising physical mechanism for tailoring light-matter interaction in a nonscattering manner \cite{R1,R2,R3,R4,anapole9}. It is generally perceived that such anapole condition requires that all of the induced electromagnetic multipole moments vanish at the same wavelength, resembling the canonical way of realizing the radiationless anapole condition \cite{anapole1,anapole2,anapole3,high_td,anapole_nl1,S_anapole1,anapole4,anapole5,anapole6,anapole_nl2,EM_anapole,anapole7,S_anapole2,S_anapole3,anapole8,anapole_LN}. In particular, the optical anapole condition corresponds to a pronounced minimum in the far-field scattering associated with highly confined electromagnetic near-field 6, 26. Such condition is quite challenge to meet since the degree of freedom for tailoring the induced electromagnetic multipolar moments of a nanoparticle is quite limited. Such a condition can also be readily fulfilled through structured light illumination \cite{S_anapole2} which generally extends the scopes of light-matter interaction from both fundamental science and application prospectives, such as high numerical aperture (NA) focusing \cite{focusing1,focusing2,focusing3}, optical computation\cite{Optical_computation}, optical data storage \cite{xu}, customized excitation of electromagnetic multipole resonances \cite{multipole1,multipole2,multipole3,multipole4,multipole5,multipole6,deng} and radiationless anapole condition \cite{S_anapole1,S_anapole2,S_anapole3}, enhancement of optical nonlinearity \cite{nonlinearity1,nonlinearity2}, optical tweezers \cite{tweezer1,tweezer2} and advanced metrology \cite{imaging1,imaging2}.

\begin{figure}[htb]
\centering
\includegraphics[width=0.95\columnwidth]{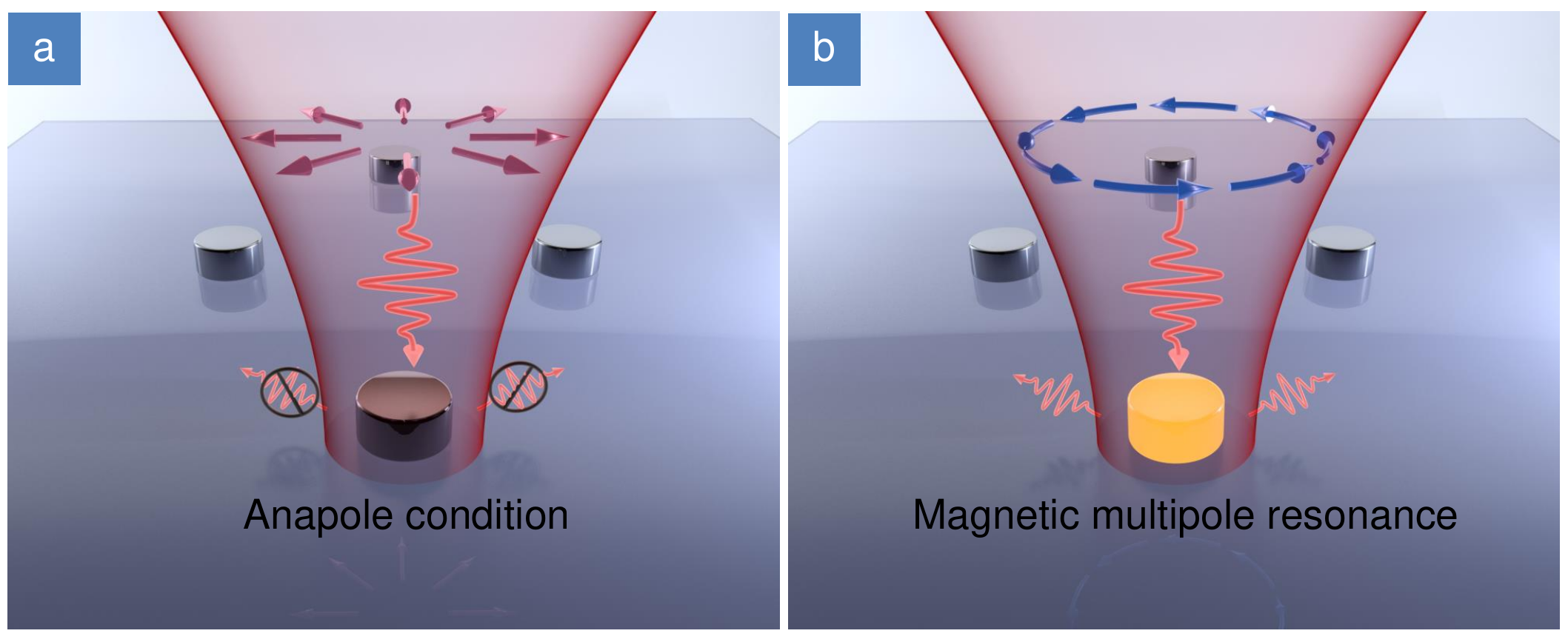}
\caption{\label{fig:fig1} 
Schematic of a reconfigurable optical antenna which supports the radiationless anapole condition hidden in a magnetic resonances at the same frequency. Tightly focused RP (a) or AP (b) beam is used to selectively realize the nonscattering and resonant scattering scenarios of the optical antenna. 
}
\end{figure} 

According to Mie theory, the total scattering power of a spherical nanoparticle excited by a plane wave is determined by both contributions of electric and magnetic multipole moments of different order $n$ \cite{book}
\begin{eqnarray}
\label{eq:eq1}
P_{total}=\frac{\pi\vert E_{i}\vert^{2}}{k\omega\mu_{0}}\sum_{n=1}^{\infty}(2n+1)(\vert a_{n}\vert^{2}+\vert b_{n}\vert^{2})
\end{eqnarray}

where $E_{i}$ , k and w are the amplitude, k-vector and angular frequency of the incident plane wave, respectively.There are numerous zeros of Mie scattering coefficients $a_{n}$ and $b_{n}$, which are corresponding to zero contribution to the total scattering from the electromagnetic multipole moments of different orders \cite{prl}. It was suggested that the electromagnetic anapole condition could be interpreted as the destructive interference between toroidal moments and Cartesian electromagnetic multipole moments \cite{anapole1,anapole2,anapole3,high_td,EM_anapole,R1,liu}. In general, the zero amplitude conditions of $a_{n}$ and $b_{n}$ are usually accompanied with other spectral overlapping electromagnetic multipole moments. As a result, such zero conditions of Mie scattering coefficients could be physically hidden in far-field scattering response. For example, zero $\vert a_{1}\vert$ can even coexist with the resonant magnetic dipole (MD) condition at the same frequency \cite{prl}. 

In this paper, we show that sophisticated tailoring of electromagnetic multipolar moments in nanoparticles is not essential for the realization of the anapole condition, which is different from previous efforts in realizing the anapole condition\cite{anapole1,anapole2,anapole3,high_td,anapole_nl1,S_anapole1,anapole4,anapole5,anapole6,anapole_nl2,EM_anapole,anapole7,S_anapole2,S_anapole3,anapole8,anapole_LN}. In particular, we experimentally demonstrate that the combination of tightly focused cylindrical vector beam (CVB) and structured Mie resonances can reveal the radiationless anapole condition which is overlapped in spectrum with a resonant state. More importantly, it subsequently enables reconfigurable optical scattering of a silicon nanoparticle with simple morphology, where the optical scattering of the nanoparticle at a specified wavelength can be switched from the radiationless anapole condition to the magnetic resonant scattering condition and vice versa, as shown in Fig. 1.

\section{Mechanism and numerical results}

\begin{figure}[htb]
\centering
\includegraphics[width=0.7\columnwidth]{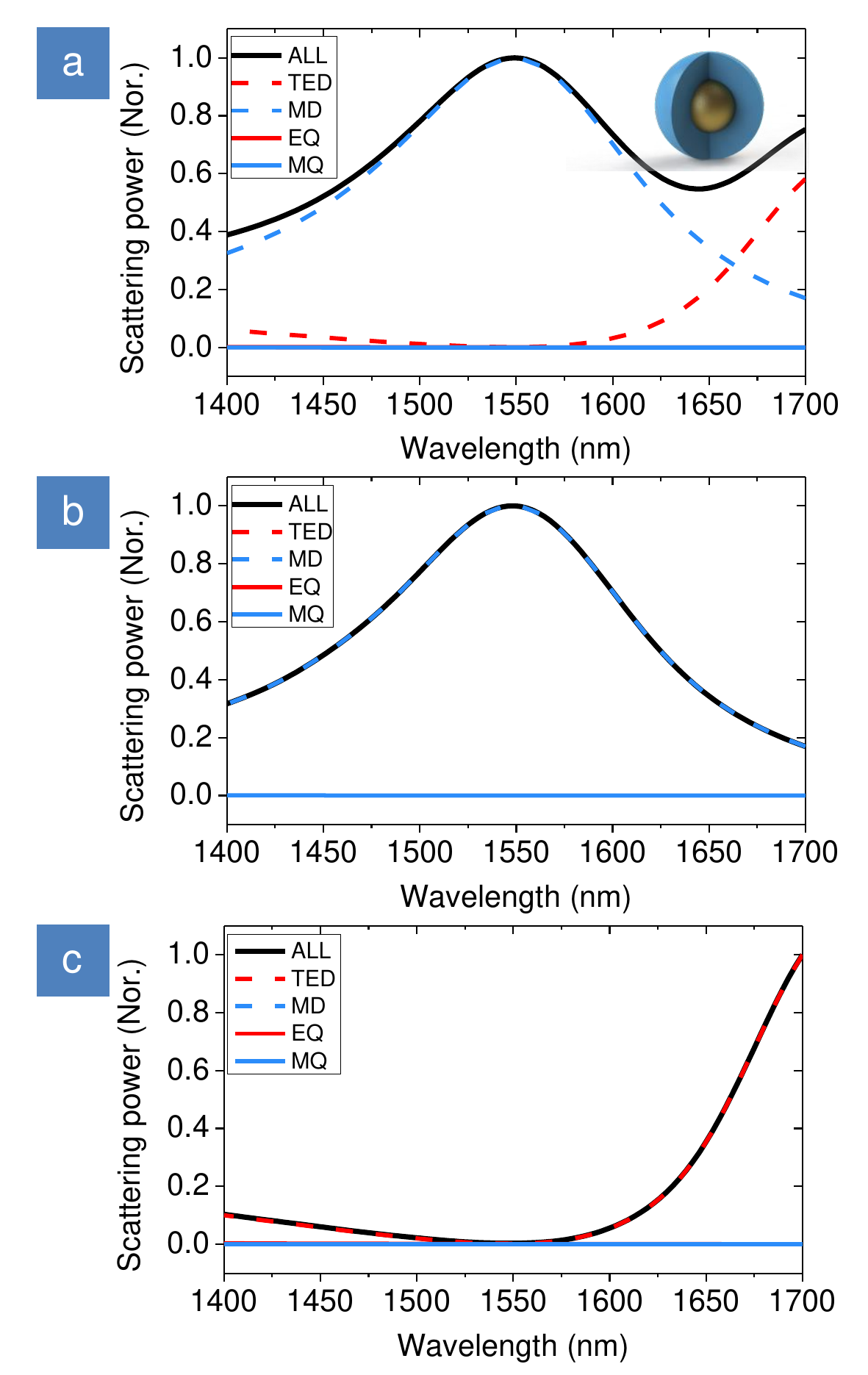}
\caption{\label{fig:fig2} 
Numerical results of electromagnetic multipolar decomposition for the normalized scattering power of a Au core/Si shell nanoparticle under the excitation by a plane wave (a), a tightly focused AP beam (b) and a tightly focused RP beam (c). The radius of Au core is 86 nm while the outer radius of Si shell is 226 nm. The NA and magnification factor of the objective lens are 0.95 and 60, respectively. TED represents the contribution of toroidal and electric dipole to total scattering. 
}
\end{figure}

The electric field of a focused CVB can be expressed in terms of the electric and magnetic multipole fields \cite{FAP_expansion}
\begin{eqnarray}
\label{eq:eq14}
E_{f}(\textbf{r})=\sum_{l=1}^\infty\sum_{-l}^{l} [p_{El}^{0}\textbf{N}_{l}^{0}(\textbf{r})+p_{Ml}^{0}\textbf{M}_{l}^{0}(\textbf{r})]
\end{eqnarray}
where $p_{El}^{0}$ and $p_{Ml}^{0}$ are the strength of the electric and magnetic multipole components, $\textbf{N}_{l}^{0}$ and $\textbf{M}_{l}^{0}$ are the vector spherical harmonics related to the electric and magnetic multipole components, respectively. For a RP beam without orbital angular momentum, the magnetic component $p_{Ml}^{0}$ is zero while the electric component $p_{El}^{0}$ is zero for a AP beam \cite{FAP_expansion}. As a result, the focused RP beam can be used to excite the electric and toroidal dipole moments because of their similarity of far-field radiation \cite{S_anapole1,S_anapole2,deng} while the focused AP beam can be used to excite the magnetic multipole modes \cite{multipole1,multipole2,multipole3,multipole4,multipole5,multipole6}. As the excited strength of an electromagnetic multipolar mode depends on both the vectorial properties of the excitation source and the eigenmodes supported by the spherical nanoparticle, therefore, the key enablers for realizing high contrast reconfigurable optical scattering include two points:1) a nanoparticle supports a pure electromagnetic multipole mode at a specified frequency, which is spectrally overlap with an anapole condition; 2) the spatial overlapping between the electromagnetic field of the nanoparticle's eigenmode and that of the tightly focused CVBs. High permittivity dielectric nanoparticles resemble a promising platform that fulfills both conditions. For a spherical Au core/Si shell nanoparticle (SN), the contributions of the spherical electric dipole (ED) can be designed to be totally suppressed at the resonant condition of MD \cite{prl}. One of the typical solution under plane wave excitation is shown in Figure 2a. If the scattering of a tightly focused AP beam is considered, it can be seen from Figure 2b that all electric multipole moments have zero contribution to the total scattering, resembling the ideal MD scattering. On the contrary, the total scattering is zero at the same frequency of the MD resonance if the excitation of a tightly focused RP beam is applied, realizing the optical anapole condition, as shown in Fig. 2c. Such a sharp contrast in optical scattering power of a nanoparticle at the same wavelength enables the realization of reconfigurable optical scattering.

\begin{figure}[htb]
\centering
\includegraphics[width=0.95\columnwidth]{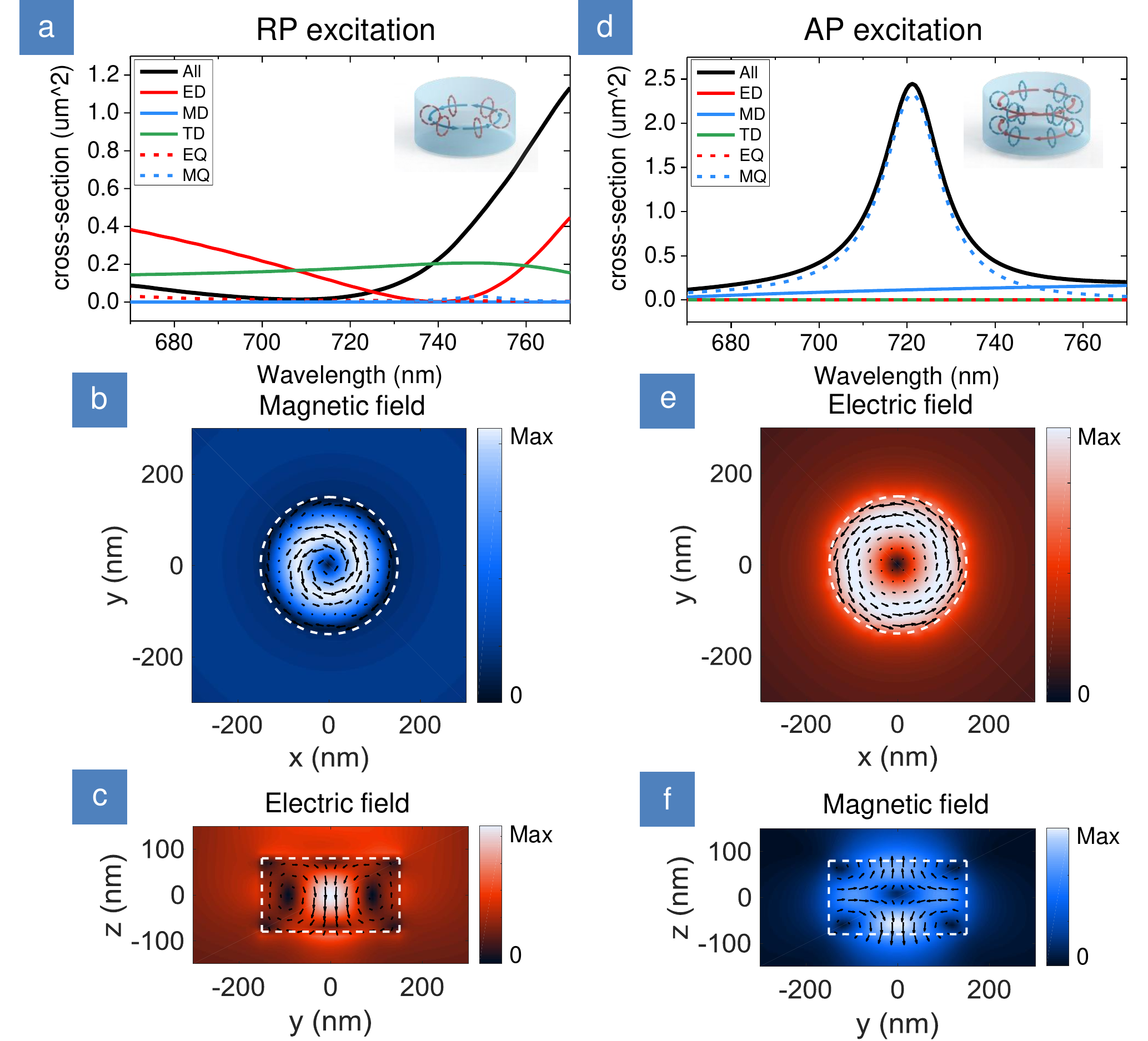}
\caption{\label{fig:fig3} 
 Cartesian electromagnetic multipolar decomposition results for the scattering power of a Si nanodisk under the excitation of a tightly focused RP beam (a) and a tightly focused AP beam (d), respectively. Multipolar moments up to quadrupole are considered. The radius of Si nanodisk is r = 150 nm while its height is h = 160 nm. The corresponding electric and magnetic field distributions at several cross sections are shown in (b), (c) and (e), (f) under the anapole condition and magnetic quadrupole (MQ) resonance, respectively. All the wavelength are 720nm. The NA and magnification factor of the objective lens are 0.95 and 60, respectively. 
}
\end{figure} 

\begin{figure}[htb]
\centering
\includegraphics[width=0.85\columnwidth]{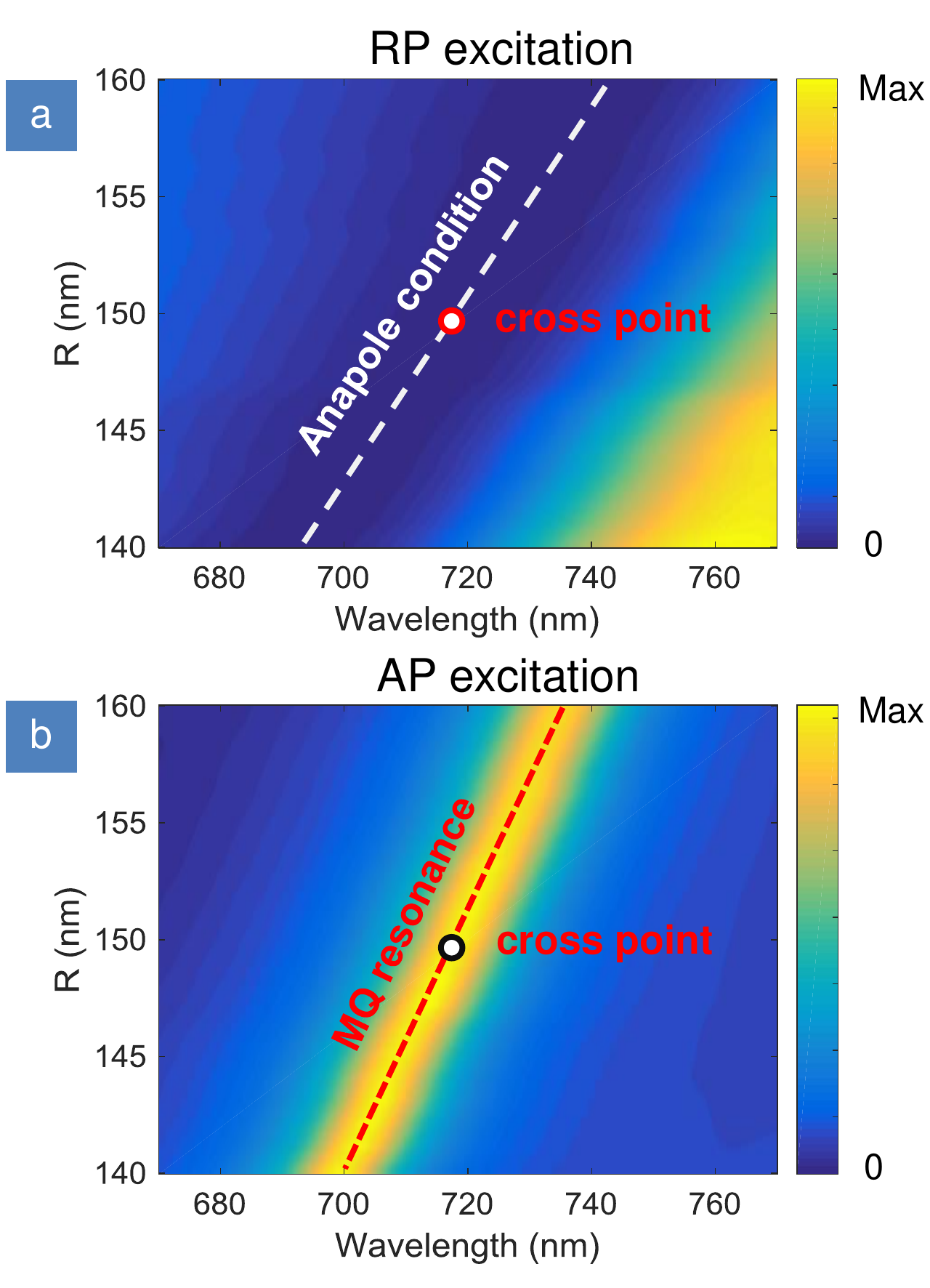}
\caption{\label{fig:fig4} 
Evolution of the total scattering spectra for the Si nanodisk with different radius under the excitation of RP (a) and AP (b) beams with fixed focusing properties. The dependence of the anapole condition and MQ resonance on the radius of the Si nanodisk are outlined by dashed lines. The NA and magnification factor of the objective lens are 0.95 and 60, respectively. 
}

\end{figure} 

In order to show the generality of this mechanism, we further consider a simple dielectric nanostructure where its magnetic quadrupole resonance overlaps with the anapole condition. Such dielectric nanostructure is much easier to fabricate in experiment than the core-shell nanoparticle. The Si nanodisk is optimized to fulfill the aforementioned enablers to realize high contrast reconfigurable optical scattering in the visible spectrum.  According to the Cartesian electromagnetic multipole decomposition results under the excitation of a focused RP beam shown in Figure 3a, the destructive interference between the Cartesian ED and toroidal dipole moments results in the anapole condition where the contributions of high order electromagnetic multipole moments are negligible. The schematic of current geometry (red) and the corresponding magnetic field (blue) are shown in the inset of Figure 3a. The calculated electromagnetic near-field distributions at different cross sections of the Si nanodisk are shown in Figure 3b and c which manifest themselves as typical signatures of near-field excitation under the anapole condition \cite{anapole3}. If the excitation source is switched to a tightly focused AP beam, the Si nanodisk is turned into the resonant scattering condition, where the dominant Cartesian electromagnetic multipole is electric quadrupole (EQ) moment. The corresponding electromagnetic near-field distributions are shown in Figure 3e and f, which validates qualitatively the dominant excitation of an MQ resonance. It should be pointed out that there is still residual induced MD moment accessed by the focused AP beam. More importantly, such high contrast reconfigurable optical scattering is tunable by simply changing the radius of the Si nanodisk, as shown in Figure 4a and b.As can be seen from this figure, the MQ resonance overlaps with the anapole condition quite well when the radius of the Si disk is changed. According to the calculated results of the Si nanodisk with different radius under the excitation of AP and RP beams, the wavelength of reconfigurable optical scattering can be effectively tuned about 50 nm, indicating sufficient tolerance for experimental demonstration.

\begin{figure*}[htb]
\centering
\includegraphics[width=0.95\columnwidth]{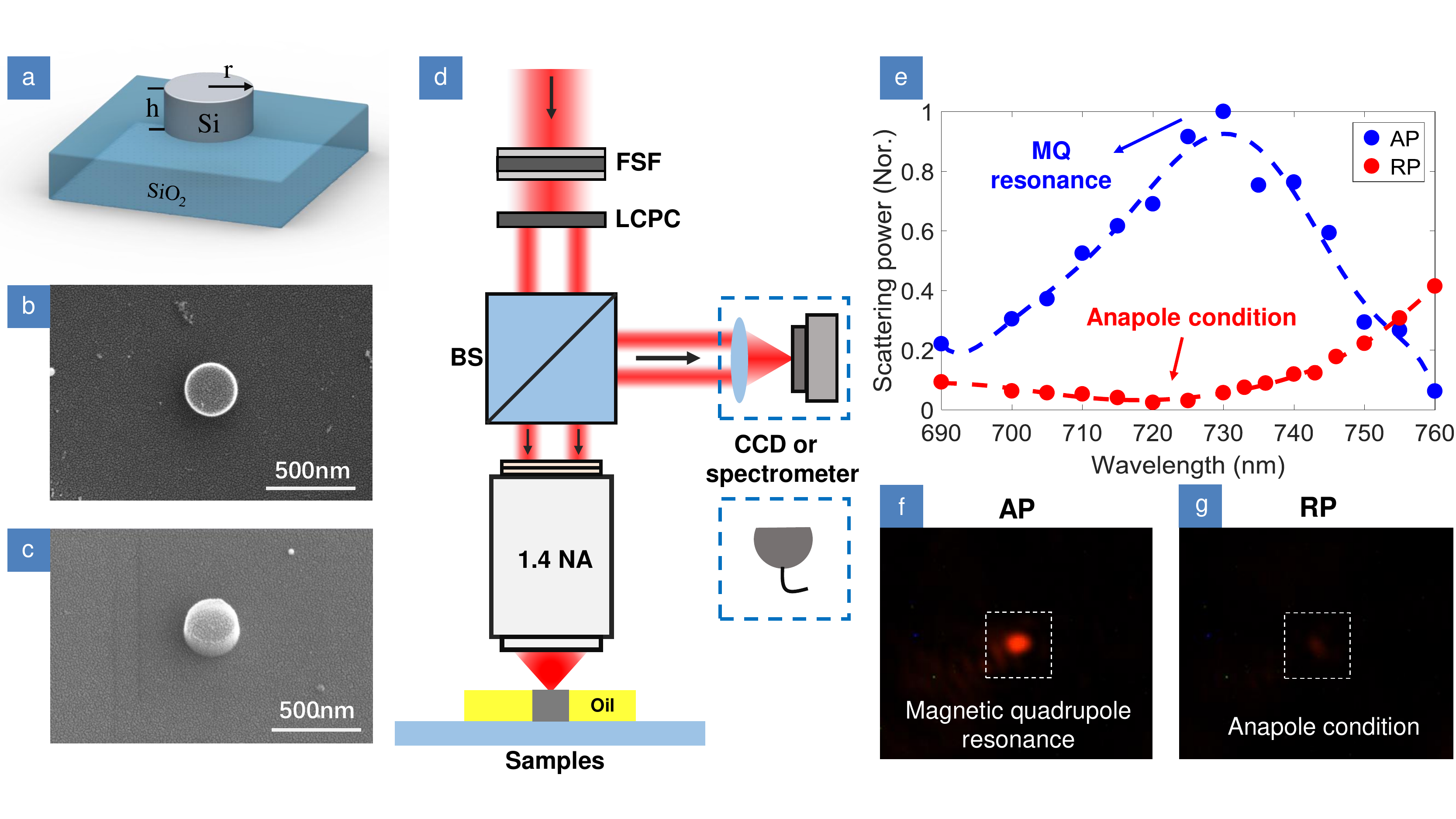}
\caption{\label{fig:fig5} 
(a) A Si nanodisk fabricated on a glass substrate, where h = 160 nm and r = 150 nm indicate its height and radius. SEM images of the Si nanodisk from top (b) and side (c) views. (d) Optical setup for the back scattering spectroscopy of CVBs. FSF, Fourier spatial filter; LCPC, liquid crystal polarization converter; BS, beam splitter; CCD, charge coupled device; NA, numerical aperture. (e) Experimental measured back scattering spectra of a Si nanodisk excited by AP (blue) and RP (red) beams under the same excitation power, respectively. The dashed lines are guide to the eye. CCD images under the MQ resonance condition excited by a focused AP (f) and the anapole condition excited by a focused RP (g). The excitation wavelengths for (f) and (g) are both 735 nm. 
}
\end{figure*} 

\section{Experimental results}

The Si nanodisk shown in Figure 3 can be readily fabricated on a glass substrate (see Supplement 1). The substrate effect only has minor effect on the results of Figure 3, as will be addressed in the following. The geometry parameters of the Si nanodisk are outlined in Figure 5a, where the radius $r$ of the Si nanodisk is 150 nm while the height $h$ is 160 nm. The top and side views of scanning electron microscopy (SEM) images of the fabricated Si nanodisk are shown in Figure 5b and c, respectively. The side view is taken by tilting the sample stage of SEM by 30 degree. As can be seen in Figure 5e, the back-scattering spectra measured by the home-built optical setup shown in Figure 5d features a resonant scattering at around 735 nm under the excitation of tightly focused AP beam. The resonance wavelength is red-shifted compared with the result of Figure 3d because of the substrate effect (see Supplement 1). However, the scattering intensity at the same wavelength is one order of magnitude smaller than the AP case when a tightly focused RP beam is used. This results quantitatively agree with the theoretical results of Figure 3. The reason for small scattering signal collected at the anapole condition might be attributed to the morphology deformation (size difference and surface roughness) in the fabricated sample utilizing mask based inductively coupled plasma reactive ion etching (see supplement 1), as shown in the inset of Fig. 5 (b) and (c). In order to further access the scatteringless anapole condition, the preparation of Si nanodisks with well-defined morphology is subject to further optimization by e-beam lithography technology \cite{6} and colloidal synthesis technology \cite{16}. Furthermore, we can confirm from the dark-field back scattering images that the radiationless anapole condition is approached under the RP excitation [see Figure 5g] while clear resonant scattering can be visualized in Figure 5f. These experimental results validate that the anapole condition can be realized without sophisticated tailoring of electromagnetic multipole moments of nanoparticle. More importantly, one can simply turn on or turn off the optical scattering of a fixed nanoparticle at the same wavelength utilizing tightly focused AP and RP beams, validating the concept of reconfigurable optical scattering for meta-optics. It is also expected that such a mechanism can be applied in the nonlinear scattering region, where distinct harmonic generation, stimulated Raman scattering and saturated scattering can be manipulated by different cylindrical vector beam.

\section{Conclusion}

In summary, we propose a new mechanism to excite the radiationless anapole condition without sophisticated manipulation of electromagnetic multipolar moments of all orders to realize superpositions of vanished moment strengths at the same wavelength. As a result, high contrast reconfigurable optical scattering utilizing the unique combination of structured light and structured Mie resonances can be realized. It means that a silicon nanoparticle whose anapole condition is hidden in an electromagnetic multipolar resonance can be selectively accessed as the radiationless state or resonant scattering state by utilizing different tightly focused CVBs. More importantly, experimental validation based on a simple Si nanodisk in the visible spectrum is provided to further consolidate the proposed reconfigurable optical scattering. Our results might provide a basic idea to realize a reconfigurable electromagnetic atom for meta-optics. By combining both degrees of freedom in structured light and structured Mie resonances, one might anticipate the possibility to realize ultrafast manipulation of optical signal without applying optical nonlinear and optomechanic effects.

\section*{Acknowledgments}
The authors acknowledge financial support from the National Key R\&D Program of China (YS2018YFB110012), National Natural Science Foundation of China (NSFC) (Grant Nos. 11674130, 91750110, 61522504 and 61975067), Guangdong Provincial Innovation and Entrepreneurship Project (Grant 2016ZT06D081), Natural Science Foundation of Guangdong Province, China (Grant Nos. 2016A030306016, 2016TQ03X981 and 2016A030308010) and Pearl River Nova Program of Guangzhou (No. 201806010040).

\section*{Conflict of interest}

The authors declare that they have no conflict of interest.

\section*{Supplementary information}
  
is available for this paper at https://doi.org/

\end{document}